\documentclass{aipproc}
\layoutstyle{6x9}
\def\be{\begin{equation}}
\def\ee{\end{equation}}

\def\bea{\begin{eqnarray}}
\def\eea{\end{eqnarray}}
\def\bml{\begin{mathletters}}
\def\blea{\begin{mathletters}\begin{eqnarray}}
\def\elea{\end{eqnarray}\end{mathletters}}
\def\cm{{\rm cm}}
\begin{document}
\title{Ultra-high-energy cosmic rays from relic topological defects}

\def\cosmologyaddress{{Institute of Cosmology,
Department of Physics and Astronomy,
Tufts University,\\
Medford, MA 02155}}
\author{Ken D. Olum}{address=\cosmologyaddress,
email={kdo@alum.mit.edu}}

\author{J. J. Blanco-Pillado}{address =\cosmologyaddress}

\begin{abstract}
It is difficult for conventional sources to accelerate cosmic ray
particles to the highest energies that have been observed.
Topological defects such as monopoles and strings overcome this
difficulty, because their natural energy scale is at or above
the observed energies.  Monopoles connected by strings are a
particularly attractive source, because they would cluster in the
galactic halo and thus explain the absence of the GZK cutoff.  Heavy
monopoles connected by light strings could last for the age of the
universe as required.  Further observations might support this model
by detection of the anisotropy due to the halo, or might refute such
models if strong clustering of arrival directions or correlations with
known astrophysical objects are confirmed.  All top-down models must
contend with recent claims that the percentage of photons among the
cosmic rays is smaller than such models predict.
\end{abstract}

\maketitle

\section*{introduction}
The observation of ultra-high-energy cosmic rays (UHECR) with energies
above $10^{20}$ eV \cite{Haya94,Bird94} is hard to explain.
First of all, is not clear that there is any site that at which
particles might be accelerated to such large energies.  Even
if such sites exist, for example in active galactic nuclei, it is hard
to explain the absence of the Greisen-Zatsepin-Kuzmin
\cite{Greisen,Zatsepin} cutoff.  Cosmic rays with energies above about
$E_{GZK} =4\times 10^{19}$ eV will interact with cosmic microwave background
photons and lose energy until they are below $E_{GZK}$.  If
the highest energy cosmic rays are produced by sources which are
homogeneous in the universe (even if some of the sources are nearby),
then there must be a large deficit in particles with $E > E_{GZK}$ as
compared to those with $E <E_{GZK}$, because the former are converted
into the latter.  Such a cutoff is not observed.

In this talk, we will try to construct a relic topological defect
model (see \cite{Pijus98} for a review of such models and others)
which addresses these difficulties.  A relic model explains the UHECR
as the decay products of some very high energy particle.  As long as
the progenitor has mass $M_Xc^2\gg10^{20}$ eV, the high energies
observed are trivially explained.  However, to explain the UHECR, the
relic must also be sufficiently long-lived to still be decaying today.
This means that the progenitor lifetime must be extremely large as
compared to the naive dimensional analysis value
$\tau\sim\hbar/(M_Xc^2)$

To solve the GZK problem, the relics must be strongly clustered near
us, and the obvious place for this is in the galactic halo.  This is
easily arranged, as long as the relic velocities are not too high to
prevent them from being captured in the gravitational potential of the
galaxy.

\section*{topological defects}
Topological defects result from misalignment of fields after symmetry
breaking transitions in the early universe.  (For a review, see
\cite{AlexBook}.) Because they are topologically stabilized, they can
persist until the present time.  If the scale of symmetry breaking is
high, for example the grand unification scale $E\sim10^{25}$ eV, then
there is no difficulty reaching the required energies for the UHECR.
However, one does need a mechanism by which the energy can be released
from the defect at the necessary rate.  One also needs an appropriate
amount of total energy stored in defects.  It must be large enough to
explain the observed UHECR flux, while not being so large as to exceed
bounds on the total density of the universe.

The dimensionality of the topological defects depends on the symmetry
that was broken to create them.  One can have monopoles (0-dimensional
defects), strings (1-dimensional defects), or domain walls
(2-dimensional defects).  If there are multiple symmetry breaking
transitions, one can have hybrid defects.  For example, a high-energy
transition can produce monopoles, and then a subsequent transition at
a lower energy can confine the flux from the monopoles into strings.
In this case, the monopoles' flux must not be the regular magnetic
field, because that is not confined today.

We can now consider the various defect types as UHECR sources.  Domain
walls are ruled out because they contribute too much total mass to the
universe.  Cosmic strings would evolve into a scaling regime, so their
total mass contribution is not too large.  However, strings move
relativistically and would not cluster in the galactic halo.
Monopoles produced with a thermal abundance would also overclose the
universe, but is possible that they were produced with a smaller
abundance during reheating or by gravitational particle creation
\cite{Kuzmin99}.  In this case, they are not ruled out, and they would
cluster in the halo.  However, it is hard to see how to get the energy
out of the monopoles.  The only real possibility is
monopole-antimonopole annihilation \cite{Hill83,Bhatta95}.
Unfortunately \cite{jjkdo99.0}, monopole-antimonopole pairs would not
be formed with sufficient density to explain the observed flux.

However, if after the monopoles have formed, a subsequent symmetry
breaking transition connects them by strings, then every monopole will
be paired with an antimonopole and there will be no problem having
sufficient annihilations \cite{jjkdo99.0}.

\section*{scenario}
We imagine a first symmetry breaking transition which gives
monopoles of mass $m_M$, and a second symmetry breaking transition
which connects them by strings of energy scale\footnote{We will
henceforth work in units where $k_B =\hbar= c = 1$.} $T_s$.  A system
of monopole and antimonopole attached by a string will oscillate with
a timescale given by the acceleration of the monopole, $\mu/m_M$, where
$\mu$ is the string tension, $\mu\sim T_s^2$.

We take the monopole not to have any unconfined flux, so the loss of
energy of the system is purely in gravitational radiation.  To produce
a long lifetime, we want small acceleration, and thus low string
tension and high monopole mass, so we take $T_s\sim100$ GeV, and
$m_M\sim10^{14}$ GeV.

To explain the observed UHECR flux we require a sufficient density of
monopoles clustered in the halo.  The minimum necessary density is
achieved when the decay lifetime is approximately the age of the universe.
In this case, for $m_M = 10^{14}$ GeV, we found the needed present
density in the universe as a whole \cite{jjkdo99.0},
\be
N_{M\bar M} > 10^{-30}/\cm^3\,.
\ee
At the time of string formation, this was
\be
n_M\sim10^{-32} T_s^3\sim10^{-18}/\cm^3\,,
\ee
which gives a typical monopole separation
\be
L_i\sim10^{-6}\cm\,.
\ee
This is much smaller than the horizon distance, $d_H \sim 3$ cm at
$T\sim 100$ GeV.

When the strings are formed they may have excitations on scales
smaller than the distance between monopoles, but these will propagate
relativistically and thus be quickly smoothed out by gravitational
radiation, leaving a straight string.  The energy stored in the string
is then $\mu L_i$.  This is smaller than the monopole mass by the
ratio
\be
{{\mu L_i}\over {m_M}} \sim 10^{-2}
\ee
so the monopoles will move non-relativistically. 

We thus have a system of monopole and antimonopole connected by a
straight string, which produces a constant force of acceleration $a
=\mu/m_M$.  The monopoles will move in elliptical orbits, but since
their thermal velocities at the time of string formation will be small
compared to the velocities they acquire during acceleration, the
motion will be nearly linear, although with enough angular momentum to
prevent the monopoles from colliding as they pass by.

To estimate the gravitational radiation rate, we will take the linear
motion, in which a half
oscillation of one monopole is parameterized by
\be
x(t) = (2 a L)^{1/2} t - {1\over 2} a t^2
\ee
for $0 < t < (8L/a)^{1/2}$. Using the 
quadrupole approximation,\footnote{The fully relativistic situation
was considered in \cite{Martin:1997cp}.} the rate of energy loss of 
the system is
\be
{{dE}\over{dt}} = {288 \over 45} G \mu^2 \left({{\mu L}\over {m_M}}\right)\,.
\ee
Now $\mu L$ is just the energy in the string, so we can write $dE/dt
=\mu dL/dt$, and integrate to get
\be
L = L_i e^{- t/\tau_g}
\ee
with
\be\label{eqn:tau}
\tau_g =  {45 \over 288} {m_M \over G \mu^2 } =
{45\over 288}{m_{pl}^2 m_M \over T_s^4}\,.
\ee

The monopoles thus move on smaller and smaller orbits, until they
annihilate, approximately when $L$ reaches the monopole core radius,
$r_M\sim m_M^{-1}$.  The system thus lives for a time about
$\tau_g\ln (L_i/r_M)$.  With $T_s\sim100$ GeV and $m_M\sim10^{14}$
GeV, Eq.\ (\ref{eqn:tau}) gives
$\tau_g\sim 10^{17}$ sec, comparable with the age of the universe.

\section*{observational consequences}

How can a model such as that presented here be verified or disproved?
Unfortunately, all models which involve topological relics or relic
particles decaying in the halo gave rise to the same observations,
dependent essentially on one unknown parameter, the mass of the
decaying primary.  Thus, the specific model of monopoles bound by
strings cannot be verified by cosmic ray observations.  However, the
low string energy scale which is necessary for long lifetimes means
that the string fields might be detected in future accelerators.

Halo relic models as a class, however, do have observable consequences.

\paragraph{Spectrum}
All relic models produce the observed cosmic ray primaries by the
decay and fragmentation of super-heavy particles (produced, in this
case, by monopole-antimonopole annihilation).  The spectrum, thus, has
little dependence on the type of defect that is decaying, but rather
results primarily from the fragmentation process.  Fragmentation of
such high-energy particles is not completely understood, but we know
that the spectrum we observe depends on the mass of the decaying
particle, and in all cases it is much harder than the steeply falling
spectrum of cosmic rays at lower energies.  Current data does not
constrain the ultra-high-energy spectrum tightly, but future experiments
\cite{Auger,OWL} should be able to validate decaying particle models
and determine the particle mass.

\paragraph{Particle type}
Fragmentation also produces a large fraction of photons, and thus a
generic prediction of relic models is that most of the observed cosmic
rays will be photons.  Identifying individual particles is difficult,
but recent studies \cite{Ave:2000nd} of large zenith angle showers
have found that no more than about 40\% of the particles can be
photons at the highest energies.  If the studies are correct, then all
relic models appear to be ruled out.

\paragraph{Anisotropy}
Because the earth is not at the center of the galactic halo, cosmic
rays coming from halo sources would be seen to somewhat higher degree
from the direction of the galactic center.  (See
\cite{MedinaTanco:1999gw} and references therein.)  The low number of
observed events, combined with the lack of an observatory in the
Southern Hemisphere where the galactic center is located, prevents a
clear confirmation or disconfirmation of this effect.  However, a
statistically insignificant anisotropy is observed of generally the
right form.  The strongest confirmation of a halo model would be to
see enhancements of the cosmic ray flux coming from the halo of M31.
Unfortunately, this also must wait for future experiments.

\paragraph{No clustering}
Any model of relic particles or monopoles will have all observed cosmic
rays coming from different sources.  Thus we would not expect arrival
directions to be clustered into multiplets, except for an effect due
to inhomogeneities in the dark matter distribution
\cite{Blasi:2000ud}.  Current claims of doublets and triplets are
consistent with dark matter inhomogeneity, but if further data yields
greater multiplets, model such as this will be ruled out.

\paragraph{No correlations with known astrophysical sources}
If the UHECR come from otherwise-invisible particles in the halo,
there should be no correlation in arrival direction with any known
object.  Such correlations have been claimed
\cite{Farrar:1998we,Virmani:2000xk}, but there is some question
\cite{Sigl:2000sn} about the correctness of these claims.

\section*{discussion}

We have argued that relic topological defects have several advantages
as sources of the observed ultra-high-energy cosmic rays.  They
naturally explain very high energies and can cluster in galactic halos
and thus explain the absence of the GZK cutoff.  However, most
topological defect models do not have the required properties.
Monopoles bound by strings seem to be a good candidate.  With heavy
monopoles and light strings, the required lifetime can be achieved,
and because there is perfect efficiency in monopole-antimonopole
binding, the required monopole density is quite small.  (Necklaces ---
monopoles connected to two strings each \cite{Berezinsky:1997td} --- also
seem like a good candidate.)

Of course, hybrid topological defects are ``exotic'', in the sense
that they involve two extra fields introduced just for this purpose.
However, since conventional mechanisms do not solve the puzzle of
UHECR origin, it seems reasonable to consider exotic models.
Unfortunately, it appears that even exotic models don't seem to be in
agreement with observation, especially the low bound on the photon
fraction from recent studies \cite{Ave:2000nd}.

\section*{acknowledgments}

We would like to thank the organizers of the 20th Texas Symposium for
an excellent conference, and Xavier Siemens, Gunter Sigl, Alex
Vilenkin, and Alan Watson for helpful conversations.  K. D. O. is
grateful for support from the Symposium and from the National Science
Foundation.  J. J. B. P. was supported in part by the Fundaci\'on
Pedro Barrie de la Maza.


\end{document}